\documentclass[sn-nature]{sn-jnl}

\usepackage{graphicx}%
\usepackage{multirow}%
\usepackage{amsmath,amssymb,amsfonts}%
\usepackage{amsthm}%
\usepackage{mathrsfs}%
\usepackage[title]{appendix}%
\usepackage{xcolor}%
\usepackage{textcomp}%
\usepackage{manyfoot}%
\usepackage{booktabs}%
\usepackage{algorithm}%
\usepackage{algorithmicx}%
\usepackage{algpseudocode}%
\usepackage{listings}%
\usepackage[separate-uncertainty=true]{siunitx}

\theoremstyle{thmstyleone}%
\theoremstyle{thmstyletwo}%

\theoremstyle{thmstylethree}%

\begin{document}


\title{Cylindrical compression of thin wires by irradiation with a Joule-class short pulse laser}

\author[1]{\fnm{Alejandro} \sur{Laso Garcia}}
\equalcont{These authors contributed equally to this work.}
\author[1]{\fnm{Long} \sur{Yang}}
\equalcont{These authors contributed equally to this work.}

\author[2]{\fnm{Victorien} \sur{Bouffetier}}
\author[2]{\fnm{Karen} \sur{Apple}}
\author[1]{\fnm{Carsten} \sur{Baehtz}}
\author[3]{\fnm{Johannes} \sur{Hagemann}}
\author[1]{\fnm{Hauke} \sur{H\"oppner}}
\author[2]{\fnm{Oliver} \sur{Humphries}}
\author[2]{\fnm{Mikhail} \sur{Mishchenko}}
\author[2]{\fnm{Motoaki} \sur{Nakatsutsumi}}
\author[1]{\fnm{Alexander} \sur{Pelka}}
\author[2]{\fnm{Thomas R.} \sur{Preston}}
\author[2]{\fnm{Lisa} \sur{Randolph}}
\author[2]{\fnm{Ulf} \sur{Zastrau}}
\author[1,4]{\fnm{Thomas E.} \sur{Cowan}}
\author*[1]{\fnm{Lingen} \sur{Huang}}\email{lingen.huang@hzdr.de}
\author*[1]{\fnm{Toma} \sur{Toncian}}\email{t.toncian@hzdr.de}

\affil[1]{\orgname{Helmholtz-Zentrum Dresden - Rossendorf}, \orgaddress{\street{Bautzner Landstraße 400}, \city{Dresden}, \postcode{01328}, \country{Germany}}}

\affil[2]{\orgname{European XFEL}, \orgaddress{\street{Holzkoppel 4}, \city{Schenefeld}, \postcode{22869}, \country{Germany}}}

\affil[3]{\orgname{Deutsches Elektronen-Synchrotron DESY}, \orgaddress{\street{Notkestraße 86}, \city{Hamburg}, \postcode{22607}, \country{Germany}}}

\affil[4]{\orgname{Technische Universität Dresden}, \city{Dresden}, \postcode{01062},\country{Germany}}

\abstract{
Equation of state measurements at Jovian or stellar conditions are currently conducted by dynamic shock compression driven by multi-kilojoule multi-beam nanosecond-duration lasers. These experiments require precise design of the target and specific tailoring of the spatial and temporal laser profiles to reach the highest pressures. At the same time, the studies are limited by the low repetition rate of the lasers. Here, we show that by the irradiation of a thin wire with single beam Joule-class short-pulse laser, a converging cylindrical shock is generated compressing the wire material to conditions relevant for the above applications. The shockwave was observed using Phase Contrast Imaging employing a hard X-ray Free Electron Laser with unprecedented temporal and spatial sensitivity. The data collected for Cu wires is in agreement with hydrodynamic simulations of an ablative shock launched by a highly-impulsive and transient resistive heating of the wire surface. The subsequent cylindrical shockwave travels towards the wire axis and is predicted to reach a compression factor of 9 and pressures above 800 Mbar. Simulations for astrophysical relevant materials underline the potential of this compression technique as a new tool for high energy density studies at high repetition rates. }

\maketitle

\section*{Introduction}

Dynamic shock compression serves as a crucial tool for creating warm and hot dense matter under extreme conditions that exist throughout the universe such as the interior of planets, supernovae, and astrophysical jets. To generate these conditions in the laboratory, a wide array of techniques are employed such as gas guns\,\cite{Fowles}, pulse power systems \cite{Deeney, XHuang}, and nanosecond high-energy laser pulses \cite{Moses, Spaeth}. Two types of shock geometries are commonly employed: planar shocks, which are prevalent in most cases, and converging shocks. Converging shocks are particularly valuable as they deliver energy to a small volume, resulting in the compression of material to exceedingly high densities and pressures. The generation of converging shocks requires precise design and facilities enabling laser irradiation with multiple beams such as OMEGA, NIF, or LMJ \cite{nora2015, boehly2011,perez2022, Doeppner2018, KrausPhysRevE, Kritcher_2016}.

X-ray Free Electron Lasers (XFEL) provide a novel platform for studying compression and shock physics. The high number of x-ray photons per pulse, low bandwidth, short temporal pulse length and high coherence make XFELs a great tool to study ultra-fast structural dynamics via a combination of techniques: x-ray diffraction, small-angle and wide-angle x-ray scattering, phase contrast imaging, x-ray absorption and emission spectroscopy, etc. The combination of the XFEL beams with high-power optical laser drivers has enabled precision measurements of extreme states of matter. The generation of high-pressure states at these facilities has been restricted to the use of high-energy (60\,J) nanosecond pulse duration lasers and recently upgraded to 100\,J. With these drivers, scientists have been able to study the equation of state and phase transitions of materials \cite{McBride2019, Kraus2016, Briggs2017}, as well as generating conditions relevant to Earth's mantle \cite{Hari_2023}, and large planet interiors \cite{Kraus2017, Hartley2018, Gleason2022, Luetgert2021}. However, the phase-space coverage is constrained in pressure range to a few Mbar due to the limited laser energy. The extension of the experimental capabilities is currently discussed by upgrade roadmaps involving coupling multi kilojoule ns-lasers at existing XFEL instruments, like the Matter at Extreme Conditions \cite{Nagler:yi5007} with the MEC-U upgrade at LCLS and the HED/HiBEF and with the HiBEF 2.0 upgrade at EuXFEL. 

On the other hand, instruments at XFELs are also equipped with short-pulse lasers delivering Joule-level energies, pulse duration of tens of femtoseconds,  and reaching intensities up to $10^{20}$\,W/cm$^2$ \cite{LasoGarcia2021, Yabuuchi2019} when focused on a sample. In this work, we show that by irradiating a thin wire with such an impulsive short-pulse laser, conditions are met where a cylindrical shock is generated and propagates towards the wire axis. At the convergence point, this shock achieves an unexpectedly high compression factor and pressure. We attribute the generation of the radial compression wave to an ablative shock created by transient resistive heating of a thin surface layer of the wire. We perform hydrodynamic simulations recovering the experimentally measured compression wave evolution. As an outlook we investigate the potential of this compression scheme for different materials relevant in astrophysical context, showing that Jovian and white dwarf conditions could be reached, enabling complementary studies to those performed at kJ-class facilities.

\begin{figure*}[htb!]

\includegraphics[width=\textwidth]{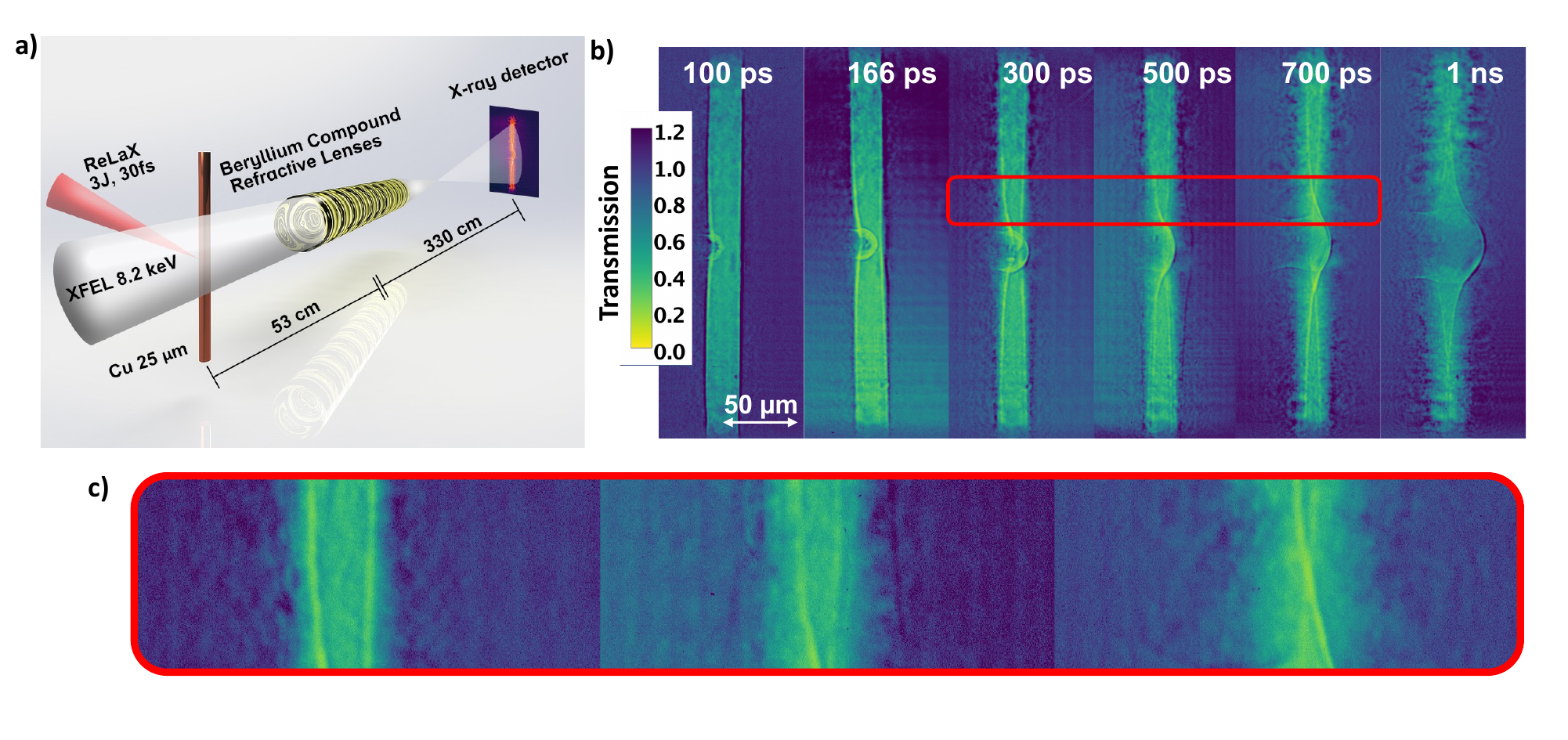}
\caption{\label{fig:fig1} a) Experimental setup of the PCI configuration used for imaging the compressed wire. The whole setup until the last 50 cm before of the detector is placed in vacuum conditions, minimizing air scattering. A slit system (not shown) is used to limit the x-ray illumination to a field of view at the sample position to  $250\times 250$\,\SI{}{\um} and minimize fringe scattering by the CRLs (300 \SI{}{\um^2} diameter) b) X-ray PCI data measured at delays from 100 to 1000\,ps after the irradiation of a \SI{25}{\um} Cu wire by a 3\,J, 30\,fs laser pulse. The color-scale gives the change in transmission compared to free beam propagation. c) Zoom into the red highlighted area of b) for improved visibility of the converging shock.}
\end{figure*}

\section*{Experimental setup for imaging the convergent shocks}
The experiment was performed at the European X-ray Free Electron Laser facility using the ReLaX laser as a relativistic plasma driver operating at the HED-HiBEF instrument \cite{Zastrau:ay5578}.  A schematic of the experiment is shown in Figure \ref{fig:fig1} a). The ReLaX laser was used at 100\,TW level, delivering laser pulses with an energy of 3\,J on target and sub 30\,fs (FWHM) pulse duration. The laser was focused employing a F/2 off-axis parabola to a spot size of approximately \SI{4}{\um} (FWHM) resulting in an average intensity of $10^{20}$\,W/cm$^2$. The 8.2\,keV x-rays generated by the SASE2 undulator were used to illuminate a square region (\SI{250}{\um})$^2$ around the ReLaX focal spot. This plane was imaged and magnified by a compound refractive lens \cite{CRLS} stack consisting of 10  beryllium lenses with a focal length of approximately 53\,cm to an imaging x-ray detector located 3.3\,m away. The detector was a GAGG scintillator imaged to an Andor Zyla CMOS camera via a 7.5$\times$ objective. The detector pixel pitch amounted to  \SI{6.5}{\um}  and after accounting the total magnification factor it results in an equivalent pixel size on target of 150\,nm/pixel. The  resolution of the imaging system was tested using calibration targets (NTT-XRESO 50HC) to resolve 500\,nm structures. It has to be noted that the resolution was limited for this experiment by the chromaticity of the SASE x-ray beam with a bandwidth of approximately 20\,eV. The temporal evolution was recorded by variation of the pump-probe delay with a precision of 200\,fs (RMS) given by the chosen temporal synchronization scheme (locked to the accelerator's radio frequency). The raw images are flat-fielded using the free beam x-ray intensity distribution (without a target) and accounting for the instrument backgrounds. The main experimental results are summarized by Figure\,\ref{fig:fig1}b)  showing Phase Contrast Images (PCI) of the wire, for different time delays ranging from 100\,ps to 1\,ns after laser irradiation. It is worth mentioning the used x-ray pulse duration $\leq 50$\,fs is much shorter than the typical few picosecond duration of laser driven x-ray backlighters used conventionally in all-optical setups \cite{Koch:98}. 
In the PCI data, the \SI{25}{\um} diameter wires are oriented vertically with the optical laser propagating from the left side and focused to the left edge of the wire in the vertical centre of the illuminated area. Besides the attenuation and phase contrast generated by the wire, the evolution of two distinct structures can be measured. During the first 300\,ps a spherical shock originating from the focal spot volume is observed similar to ref. \cite{Santos_2017}. At 300\,ps after the laser irradiation, this shock has already propagated through the wire.  At the same time one can follow a second nearly-cylindrical shock moving radially inward towards the wire axis originating from both left and right edges of the wire. The velocity of this shock is decreasing with increasing distance from the ReLaX focus.  At  $500-1000$\,ps a convergence of the shock close to the axis of the wire can be seen. Simultaneous to the inward radial motion, a radial expansion resulting into a smoothing of the wire edge is observed. This effect is attributed to wire plasma expansion.  To quantify the evolution associated with the radially converging structure we have analyzed lineouts at \SI{42} and \SI{100}{\um}  from the focal position of the laser. 
   
We calculate the velocity of the shock by measuring the distance travelled by the shockwave between the time delays of 300\,ps  and 500\,ps.  These time delays lie within a constant shock velocity region, avoiding the initial deceleration as well as the final acceleration, as predicted by simulations and explained by the next section. An average velocity  of $14.3 \pm 1.3$\,km/s is observed \SI{42}{\um} close  to the laser focus, and  of $10.5 \pm 1.3$\,km/s at \SI{100}{\um}. Additionally we evaluated the shock front speed emerging from the laser focus. Here the velocity of the front decreases from 180\,km/s at 20\,ps to 50\,km/s at 300\,ps the time of the shock release.

\section*{Origin of the shocks unraveled by numerical simulations}
 In this section the origin of the observed cylindrical converging shocks is investigated. It is instructive to start by looking at the details of the laser-wire interaction.  Particle in Cell simulations are commonly used for simulation of laser-matter interaction, as predictive tools giving detailed insights into plasma evolution at the time of laser interaction and shortly after (at ps timescales). Performing a simulation with a wire as target one would observe processes similar to Refs. \cite{Quinn2009, Bouffechoux10}: first the generation of a highly energetic electron population by the direct interaction of the impulsive laser pulse with the wire material that will propagate longitudinally through the wire, leading for example to generation of a plasma sheath with strong electric fields that accelerate ions \cite{Wilks01}.  Simultaneously, part of the hot electron population will move with close to speed of light transversely away from the focal spot.  The hot electrons are electrostatically trapped close to the wire surface, and in their wake the wire surface is ionized.  To achieve charge balance a return current along the surface is established reaching current densities of $10^{13} $A/cm$^2$.  This currents encompass the whole wire surface up to a skin depth.  Proton imaging techniques (employing laser accelerated protons) have been successfully applied to measuring the sheath fields and thus the dynamics of the associated hot electron transport and subsequent return current \cite{Borghesi2009,Quinn2009,10.1063/1.3262630,Borghesi2005a}, both in simple wire targets but also in foils \cite{Romagnani2005}, or more complex targets \cite{Toncian2006}, and also to understand late time dynamics of the hereby measured instabilities \cite{Quinn2012}. 

While the lifetime of such currents was estimated by Quinn {\it et al.} to be 20\,ps \cite{Quinn2009},  one order of magnitude longer than the ps duration of the laser pulse used, recent theoretical work has been focused on laser pulses matched to our experiment of several tens fs \cite{yang2023dynamic}.   This study reveals that the surface return current itself lasts in the order of 100\,fs for a 30\,fs laser pulse. The surface return current has two effects on the wire target which can lead to compression, the magnetic compression and Joule heating and the associated ablation. It is also shown that the return current magnitude scales with the wire radius. For a \SI{25}{\um} diameter copper wire, the peak of the surface return current density is predicted to be in the range of 40-\SI{110}{kA/\um^2}, with a current strength decaying further away from the laser focus. With this surface return current, the $\textbf{J} \times \textbf{B}$ force can be neglected while Joule heating arises as the mechanism to drive convergent shocks. The  resulting electron temperature distribution is estimated  using the electron energy equation \cite{Beg2004,yang2023dynamic},
\begin{equation}
\frac{3}{2}n_e\frac{\partial T_e}{\partial t}=\frac{\partial}{\partial r}(K_{T_e}\frac{\partial T_e}{\partial r})+\frac{j_h(r)^2}{\sigma_{T_e}},\label{eq:heat_transfer}
\end{equation}
where $K_{T_e}$ is the thermal conductivity of cold electrons, $\sigma_{T_e}$ is the electric conductivity and $j_h(r)$ is the surface return current distribution in the radial direction. The equation's right side accounts for heat diffusion in the radial direction and the Joule heating due to the surface return current. The electron resistivity model and the heat diffusion coefficient are functions of temperature and density and can be extracted from the SESAME equation of state \cite{johnson1994sesame}. The calculated electron temperature with this equation peaks at 140-320\,eV for the return current range of 40-\SI{110}{kA/\um^2} and is distributed within a \SI{0.1}{\um} skin depth layer. 
This temperature distribution is used as the initial condition for hydrodynamic simulation to investigate plasma evolution until 1\,ns.  These simulations with the FLASH code \cite{fryxell2000flash,dubey2009extensible} solve in a 1D cylindrical symmetry the one fluid, two-species (ion and electron) and two-temperatures hydrodynamic equations with copper SESAME equation of state\cite{johnson1994sesame}. The ion temperature and electron temperature are assumed to be equal due to the high collision rates between these.

\begin{figure}[htbp!]
\includegraphics[width=\linewidth]{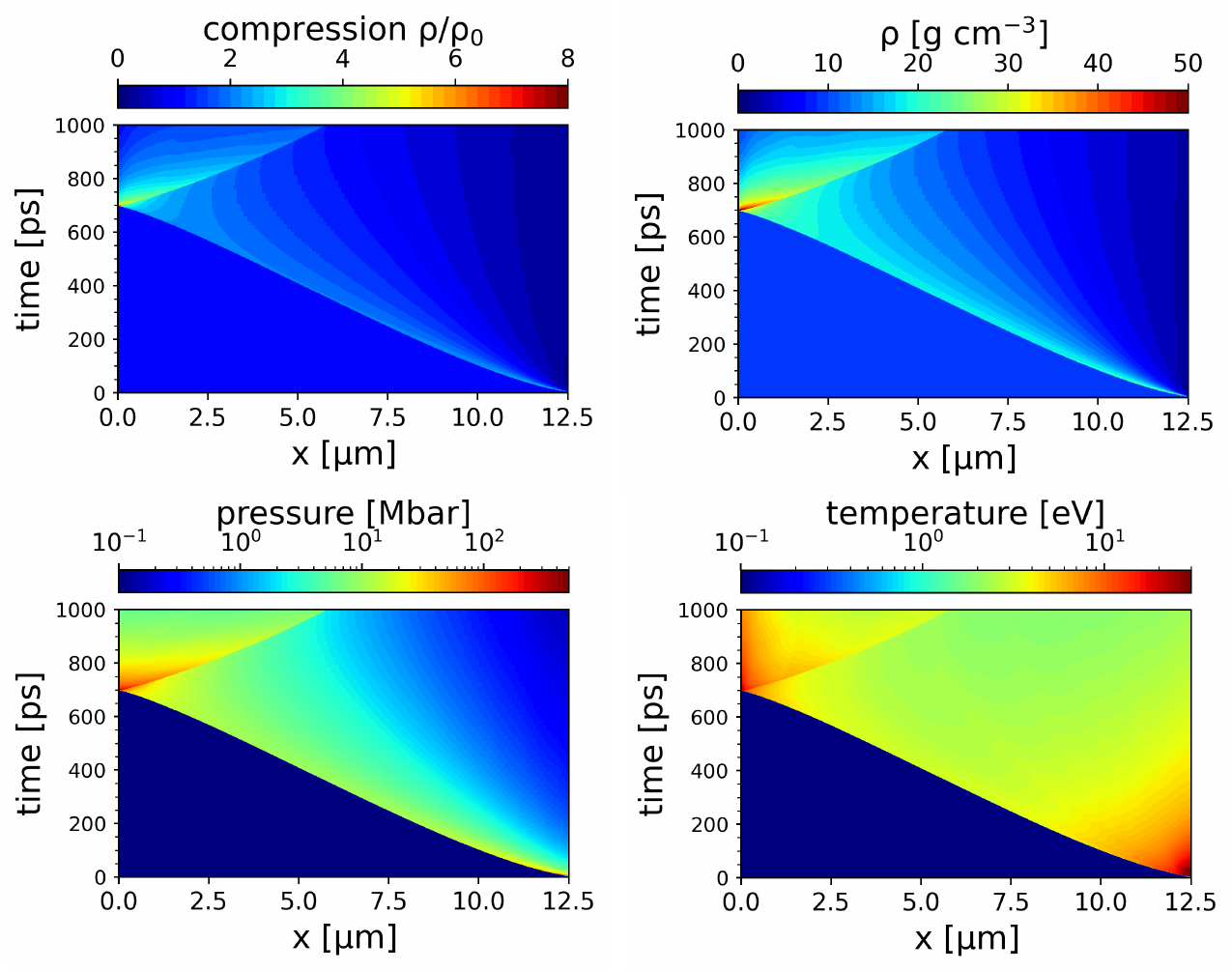}
\caption{\label{fig:fig2} a) Compression factor evolution for several time steps. b) density, c) pressure and d) temperature evolution  for shock driven by a surface temperature of 250\,eV  obtained from hydrodynamic simulations using the SESAME equation of state.}
\label{figure4}
\end{figure}

Figure \ref{fig:fig2} shows the result of the simulation of the shock dynamic for a \SI{25}{\um} copper wire under the experimental laser irradiation conditions, giving the temporal and spatial evolution of density, compression (density normalized to initial density), pressure and temperature. The initial condition used is a peak temperature of 250\,eV with an exponential decay depth and a decay constant $\tau = 0.067$\,\SI{}{\um}. The shock is formed within the first 5\,ps due to the ablation pressure. It starts at the surface with a compression factor of ~2.8 with respect to cold copper and a peak pressure of 111 Mbar. It travels towards the wire axis with a starting velocity of 44\,km/s. At a time of 200\,ps the shockwave has decelerated to 15\,km/s and the temperature of the shock front reaches 22\,eV, while the pressure decreases to 12\,Mbar. The peak compression factor at this point is 2. Between 200\,ps and 650\,ps the shock propagates with constant conditions and the shock front moves from \SI{8.1}{\um} away from the wire axis down to \SI{1.1}{\um}. After this point, the shock gains velocity until it converges at the wire axis reaching compression factors of 9, corresponding to densities of \SI{80.6}{g/cm^{3}}, a temperature of 38\,eV and a pressure of 830\,Mbar.

\begin{figure}[htbp!]
\includegraphics[width=\linewidth]{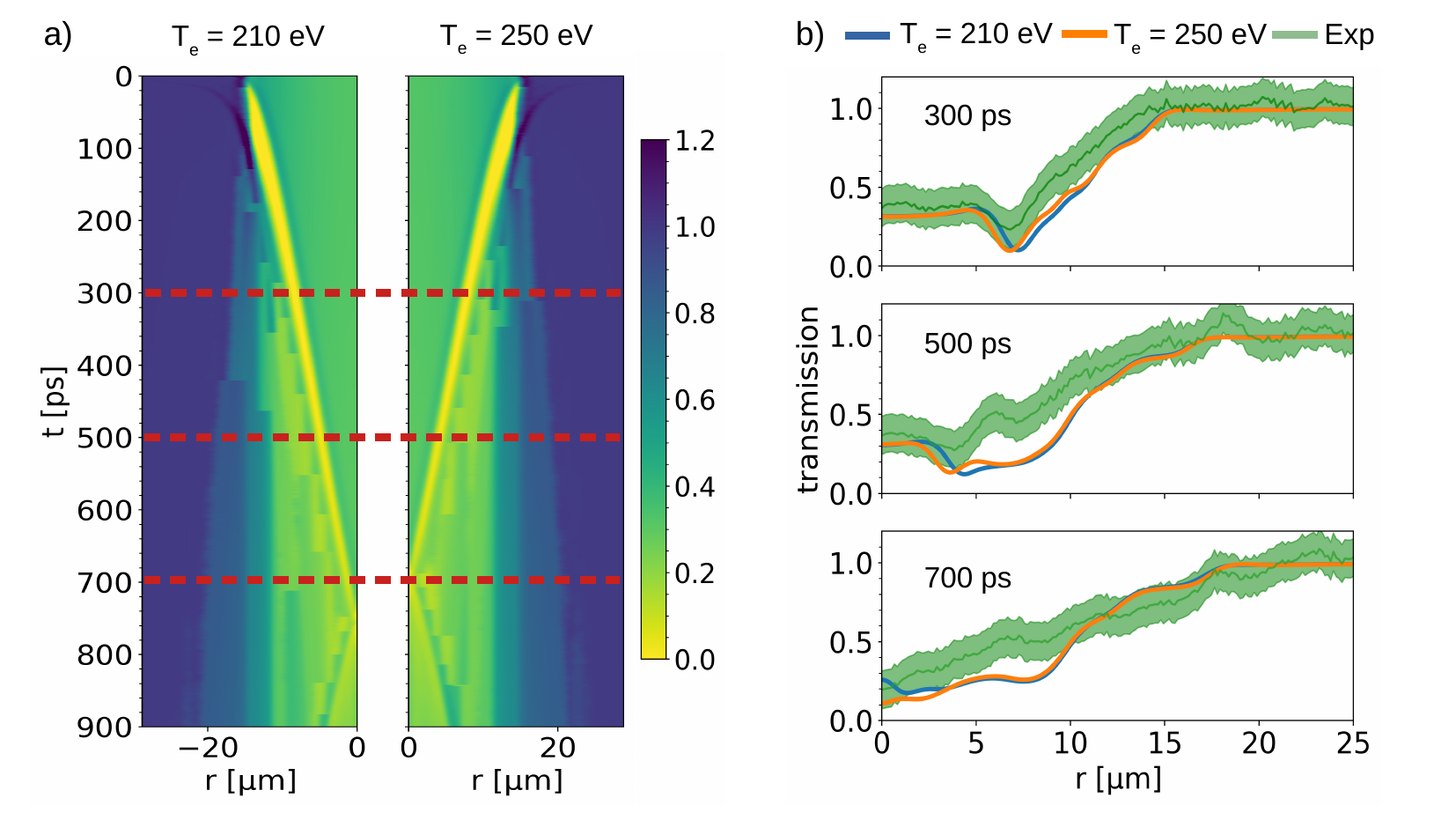}
\caption{\label{fig:fig3} a) simulated PCI profiles using the hydrodynamic simulation data for initial temperature of 210 and 250\,eV. b) Comparison of experimental and simulated data 300\,ps, 500\,ps, and 700\,ps for 210\,eV and 250\,eV initial temperature. The matching of the experimental observed convergence at 700\,ps occurs for the 250 eV case.  } 
\end{figure}

\section*{Comparison between experimental and synthetic imaging data}
Using the radial density profile from the hydrodynamic simulations the expected x-ray PCI profiles were calculated and compared with experimental results. Using a forward Abel transform of the density profile  the projected mass density is obtained as probed by the x-ray in the experimental geometry. The intensity at the detector plane can be calculated via the Transport of Intensity Equation \cite{Teague:82}, as:

\begin{equation}
    I(x_1,z=z_{1}) = I(x, z=0) \left(1 + \frac{z_{1}}{k} \nabla^{2} \Phi(x, z=0)\right)^{-1} 
\end{equation}

where $I(x_1,z=z_{1})$ is the intensity at the detector located at a propagation distance $z_1$, $I(x, z=0)$ is the intensity at contact, that is directly at the exit plane of the target and given by the attenuation of the x-rays by the target, $k$ is the x-ray wave-vector and $\Phi(x, z=0)$ is the phase shift at contact. The PCI data measured with an undriven wire was used to characterize the propagation distance, resulting in an equivalent plane located at a distance $z_1 = 6$\,mm after the target that is imaged by the CRLs onto the detector.
Figure \ref{fig:fig3} a) shows the temporal evolution of the forward calculated PCI pattern up to 1\,ns for two initial temperatures, 210\,eV and  250\,eV. The convergence time is 758\,ps for the 210\,eV simulation and 698\,ps for 250\,eV. The resulting profiles at time-steps 300, 500 and 700\,ps are compared in \ref{fig:fig3} b) with experimental data \SI{42}{\um} away from the laser focus. The synthetic profiles reproduce  quantitatively and qualitatively the experimentally observed PCI patterns with features such as position of the inward propagating  shock-front and outward beam refraction. In this context it is important to consider the effect of  temperature-induced opacity changes.  Using the TOPS / ATOMIC database  \cite{tops} it was confirmed that the values for the mass attenuation coefficient for a temperature of 22\,eV differ on percent level from those of cold copper material, thus it can be neglected for the rest of the analysis.%

We have selected the experimental data at 300\,ps and 500\,ps delay and further analyzed the PCI profiles at \SI{42}{\um} away from the laser focus to extract the shock parameters. As the shock acceleration is minimal between these delays,  the  uncertainty in the velocity estimation and its effect on the shock pressure is reduced. The generalized Paganin method \cite{Paganin_2020} was applied to calculate the intensity at the target plane. Here the assumption is that all the intensity variation is due to absorption in the wire. An inverse Abel transform is used to extract the radial mass attenuation and consequently  the density profile. Finally the Rankine-Hugoniot equation is used to calculate the shock pressure using the experimental shock velocity and density, obtaining a value $p = 11.0 ^{+4.0} _{-2.6}$\,Mbar for the 300\,ps and $p = 10.0 ^{+3.9} _{-2.6}$\,Mbar for 500\,ps. The values extracted from hydrodynamic simulations are 10.5\,Mbar and 10.6\,Mbar, in close agreement with the experimental value.

\begin{figure}[tbp!]
\includegraphics[width=1\linewidth]{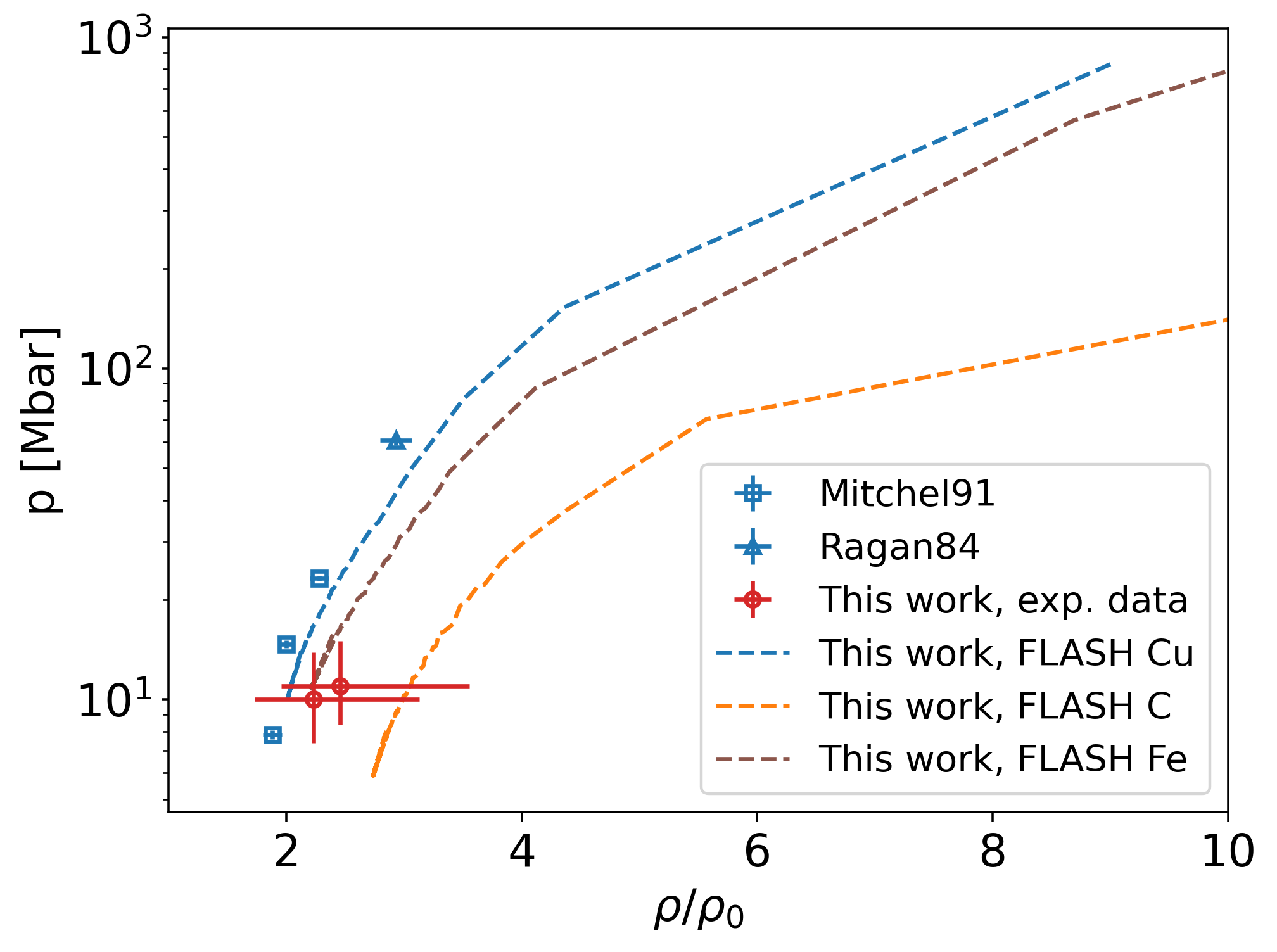}
\caption{\label{fig:fig4} Shock states for different materials: the red dots represent the pressure extracted at delays of 300\,ps and 500\,ps for copper by this work, the squares and triangle corresponds to previous published results \cite{Mitchell, Ragan}.  The blue dashed line are states reached for Cu according to hydrodynamic simulations, the brown and orange lines are Fe and C simulations.} 
\end{figure}

\section*{Scaling for various materials}
A comparison of this value with previous published results for copper is shown in Figure \,\ref{fig:fig4} together with the predicted states achieved in our experiment according to the FLASH simulations. Furthermore, Figure \ref{fig:fig4} shows simulation predictions for carbon and iron as representative materials in the context of astrophysical research. 
Using the same return current conditions and target diameter as for the Cu wires, the simulations predict pressures up to 790\,Mbar for iron and up to 400\,Mbar for carbon. The temperatures in this cases range up to 32\,eV and up to 16\,eV for iron and carbon, respectively at the time of convergence. The carbon states are comparable to the ones expected in Jovian worlds as well as exoplanets \cite{planetpaper}, showing the potential of this platform for planetary interior research. The iron states are in the range of the stellar conditions for white dwarf envelopes as shown in MJ experiments at NIF \cite{Kritcher2020}.  While further investigation are beyond the scope of this paper, the consideration of higher dimensionality of the compression  is paramount for a fully quantitative prediction of the compression capabilities. First 2D simulations assuming a 33\% drop of the initial temperature between the front and back surface of the wire show that the maximal density would decrease by 25\% compared to the case of ideal compression, demonstrating the robustness of the process and potential of this method as platform for HED studies.

\section*{Summary}

In summary, we have demonstrated the capabilities of a Joule-class laser irradiating  thin wire targets to generate extreme pressure states relevant to astrophysical studies. We have shown how  the state can be characterized via imaging techniques exploiting the ultra-short duration and high brilliance of an XFEL beam. In particular, converging cylindrical shocks in copper with pressures up to 11\,Mbar have been measured, with simulations predicting pressures  of to 830\,Mbar at convergence, supported by the excellent quantitative agreement between experimental and forward calculated data.
This method of shock-generation paves the way to performing astrophysical experiments in the laboratory providing large statistics thanks to the high repetition rate of the lasers (shot per minute) and involving simple and ubiquitous targets.


\section*{Acknowledgments}

We acknowledge the European XFEL in Schenefeld, Germany, for provision of X-ray free electron laser beam time at the Scientific Instrument HED (High Energy Density Science) and would like to thank the staff for their assistance. The authors are indebted to the HIBEF user consortium for the provision of instrumentation and staff that enabled this experiment. FLASH  was developed in part by the DOE NNSA- and DOE Office of Science-supported Flash Center for Computational Science at the University of Chicago and the University of Rochester.

\subsection*{Authors contributions}
A. L. G. and T. T. conceptualized the experiment. L. Y. and L. H. developed the theoretical framework. A. L. G., V. B., K. A., C. B., H. H., O. H., M. M., M. N., A. P., T. R. P., L. R. and T. T. performed the experiment. A. L. G., L. Y., T. E. C., J. H. and T .T. analyzed the data. A. L. G., L. Y., L. H., T. E. C. and T. T. wrote the original manuscript draft. All authors reviewed and edited the manuscript. L. H. and T. T. supervised the project.

\subsection*{Competing interests}
The authors declare no competing interests.

\subsection*{Data availability}
Data recorded for the experiment at the European XFEL are available at \cite{data}.


    
\begin{appendices}

\section*{Methods}
\subsection*{X-ray setup}
\begin{figure}[ht!]
\includegraphics[width=1\linewidth]{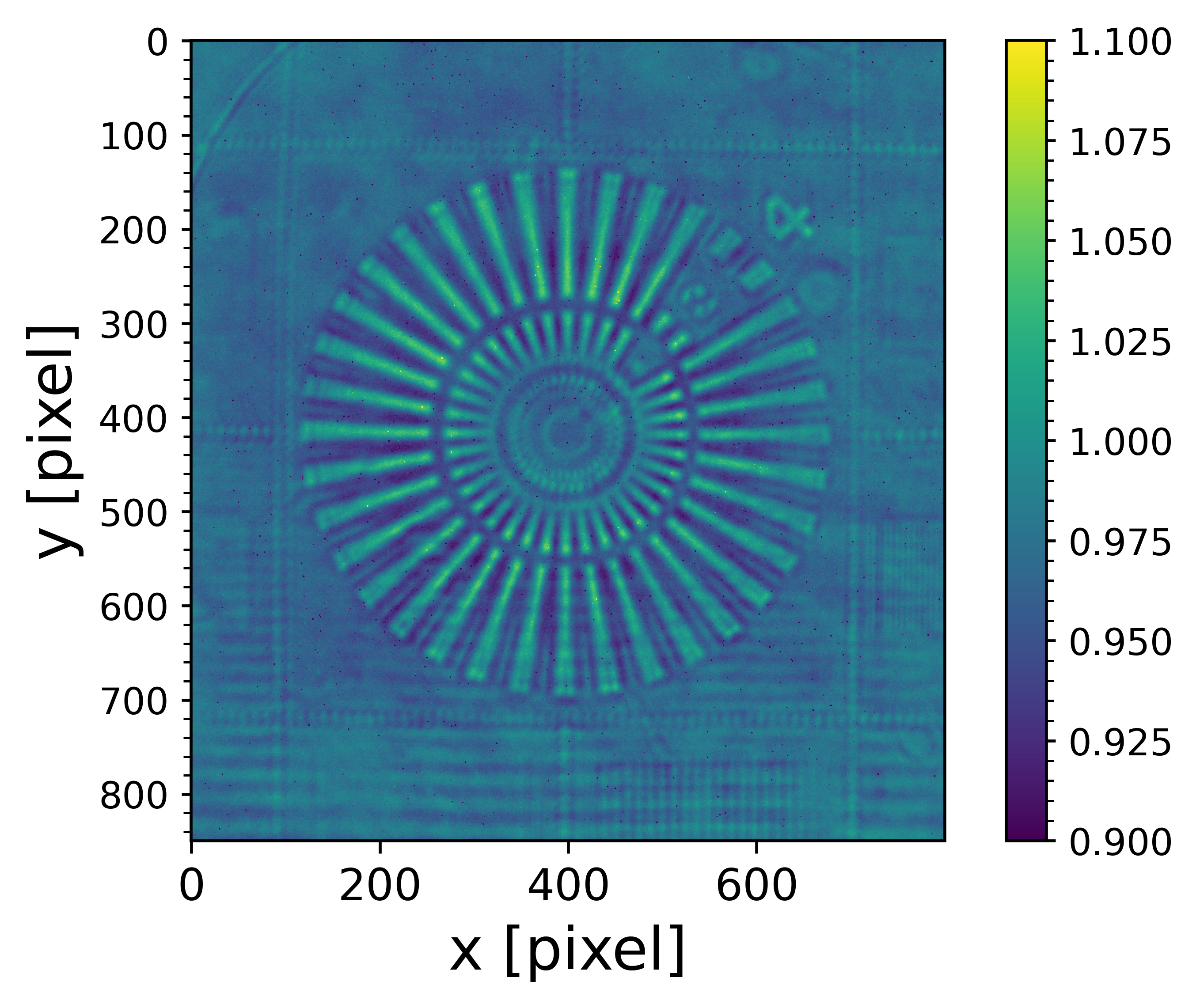}
\caption{X-ray imaging of a Siemens star calibration sample.}
\label{fig:s1}
\end{figure}

The x-ray beam was characterized in energy via the elastic scattering of the beam on a YAG scintillator. The elastic signal was measured via a von Hamos x-ray spectrometer \cite{Preston_2020}, which was previously calibrated via copper K$_{\alpha}$ emission. The energy was determined to be 8.2\,keV. The pulse energy of \SI{600}{\uJ} on average was measured via an x-ray gas monitor (XGM) in the x-ray tunnel. A x-ray lens configuration was chosen that resulted in a pencil-like beam at the target chamber (vacuum  in $10^{-5}$ mbar range).The beam size was measured with a YAG scintillator at the pulse-arrival-monitor located 9.5\,m before the target-chamber-center (TCC) and at 3.3\,m after TCC with the same detector used for PCI measurements. The beam size at TCC was interpolated between those two points.
The compound refractive lenses stack consisted of 10 Be lenses, with a radius of curvature of \SI{50}{\um} manufactured  by RXOptics Germany with a wed thickness of \SI{50}{\um} for each lens.
The resolution of the system was characterized by imaging of a Siemens star test target, NTT-XRESO-50HC. The resolution target is made of tantalum with a thickness of \SI{500}{nm}. The imaging of the target is shown in Figure~\ref{fig:s1}.

\subsection*{Optical laser setup}
The ReLaX laser was used at 100\,TW energy level, delivering 3\,J of energy on target. The pulse duration was optimized using a self referencing spectral interferometer WIZZLER 800 by Fastlite and was regularly checked for best compression by a second harmonic generator autocorrelator. The pulse intensity contrast was measured to be within the specs presented by \cite{LasoGarcia2021}. The focal spot quality was monitored and optimized by using a 20x APO PLAN microscope objective. The spatial phase was optimized by an adaptive deformable mirror coupled to a wavefront sensor. The synchronization between optical and x-ray laser was measured by spatial photon arrival monitor techniques \cite{Kirkwood2019}.
\subsection*{Data processing}
Each of the x-ray images taken was flatfielded according to the following procedure: first the detector background is subtracted by subtracting the average of 313 empty frames. Then the image is normalized by the pulse energy measured by the XGM. Next, the scattering pattern generated by the slits on the shot of interest is compared to an ensemble of scattering patterns taken without target. The slit scattering pattern is sensitive to the x-ray intensity and the beam pointing. A chi-square minimization is used to find the best match. The normalized on-shot image is divided by the normalized free-beam best match. Finally, any residual intensity variation due to imperfect match is locally corrected by fitting a third order polynomial to an area 200 pixels around the wire shadow and dividing by the fit result. The final uncertainty of the flatfield is obtained from the peak-to-valley transmission variation associated to the transmission baseline. This results in an uncertainty of $s(T) = 0.12$.

\begin{figure}[ht!]
\includegraphics[width=1\linewidth]{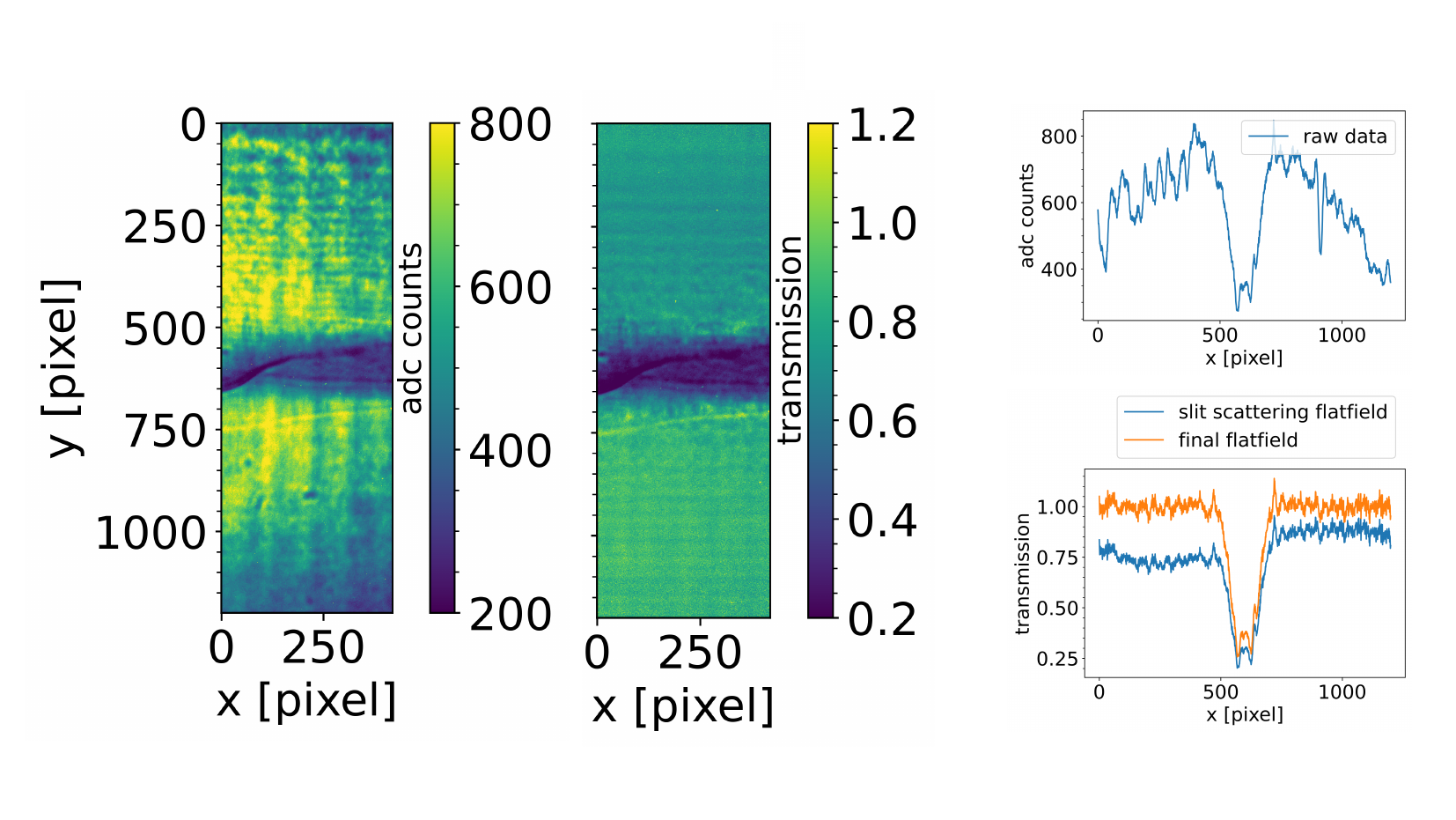}
\caption{Experimental data flatfielding steps. a) The raw data as measured by the detector. b) Flatfielded data reducing the slit scattering. c) Lineout of the raw data at a distance \SI{42}{\um} from the laser interaction point. d) Lineout of the flatfield by the slit scattering and the final flatfielding by fitting the intensity outside the wire shadow.}
\end{figure}

\subsection*{Hydrodynamic simulations}
The FLASH code as version 4.6.2, developed by the University of Rochester, was employed for hydrodynamic simulations. These simulations utilized a 1D cylindrical symmetry geometry. The total simulation box is \SI{50}{\um} in size, with the wire target material occupying \SI{12.5}{\um} and the rest containing vacuum. This vacuum is filled with low-density hydrogen at a density of $10^{-5}$\,\SI{}{g cm^{-2}} and a temperature of 1\,eV. The target material varies depending on the case and includes copper, iron, or carbon. Each target region contains solid targets at their respective solid densities. The initial temperature distribution of the target is determined using the electron energy equation 1. Based on the return current scaling theory, magnetic compression driven by the $\textbf{J} \times \textbf{B}$ force is disregarded, setting the initial fluid velocity to zero. The boundary conditions use a reflective boundary condition for the symmetry axis and free-space for the vacuum. A self-adaptive mesh grid and derived material properties from the corresponding SESAME equation of state was used. 

\subsection*{Synthetic PCI data}
The density output of the hydrodynamic simulations is used to calculate the synthetic PCI intensity $I(x,z)$ with x the transverse distance from the axis of the wire relative to the x-ray propagation direction and z the distance from the wire  along the x-ray propagation distance. In a first step the Abel transform is evaluated $$\Gamma(x)=2\int_x^\infty \rho(r)r/\sqrt{r^2-x^2}dr,$$ to obtain  $\Gamma$ the total mass density projected along the line of sight x using $\rho(r)$  the 1D mass density in cylindrical coordinates. From $\Gamma(x)$ the absorption mass attenuation coefficient $\mu(x)$ and the phase $\Phi(x,z=0)$ are calculated at the exit plane of the wire. The change of PCI intensity after a propagation distance z is given by $$I(x,z)=I_0 e^{-\mu(x)}\left( 1+z/k\nabla^2 \Phi(x,z=0)\right)^{-1},$$ with $I_0$ the original intensity, k the wave number of the x-ray.  In a last step we are considering both the experimental resolution and also the bandwidth of the SASE x-ray beam. The limitations of the resolution are simulated by applying a Butterworth filter with a cut-off frequency of \SI{0.00057}{nm^{-1}}  corresponding to the Nyquist frequency of the detector imaging system. The SASE bandwidth effect is calculated by approximating the spectral distribution as a Gaussian with 20\,eV FWHM and sampling 40 wavelengths $\omega$. For each of these wavelengths a corrected z (compared to the 6 mm)  is used to obtain $I(x,z(\omega))$. The final PCI intensity is integrated from the $I(\omega)$ with weights given by the spectral distribution.

\subsection*{Shock density reconstruction}
The procedure to reconstruct the density from experimental PCI profiles is the direct inversion of the procedure to produce the synthetic PCI data. The Paganin method is used to invert the TIE equation and obtain the intensity profile at target contact. The intensity at contact, $I(x,y,z=0)$ relates to the measured one, $I(x,y,z)$ as $$I(x,y,z=0) = - log_e \left(\mathcal{F}^{-1}\frac{{\mathcal{F}[I(x,y,z)/I_0]}}{1-2z\frac{\delta}{\mu d^{2}} (cos(d k_x) + cos(d k_y) - 2)}\right)$$ where $\mu$ is the mass attenuation coefficient, $\delta$ is the real part of the index of refraction, $d$ is the pixel size on target and $k_{x,y}$ are the spatial frequencies at the detector plane, and $\mathcal{F}$ the Fourier transform.
The intensity at contact is then directly related to the mass attenuation coefficient. The attenuation as a function of radius is calculated via the inverse Abel transform of the intensity. Finally, division by the mass attenuation coefficient returns the radial density profile. And estimate of the uncertainty of the abel reconstruction was obtained by shifting the center of rotation by a distance of 500\,nm around the nominal axis to account for the finite resolution. This procedure was done for the experimental intensity profile as well as the upper and lower boundaries. The final peak density and uncertainty is calculated as the average and standard deviation of the peak density of all profiles. The peak density for the 300\,ps data point is $\rho = 22 ^{+10} _{-5.0}$\,$\mathrm{g cm^{-3}}$, for the 500\,ps $\rho = 20 ^{+8.4} _{-4.5}$\,$\mathrm{g cm^{-3}}$.

\subsection*{Calculation of the shock pressure}
The pressure was extracted from the experimental data via the Rankine-Hugoniot relation. The pressure can be expressed as: $$p = p_0 + u_s^2 \rho_0 \left( 1 - \frac{\rho_0}{\rho_s} \right)$$ where $p$ is the shock pressure, $p_0$ is the pressure of the matter in front of the shock, $u_s$ is the shock velocity, $rho_0$ is the uncompressed density and $\rho$ is the compressed density. In our case, the pressure in front of the shock is ambient pressure. The shock velocity can be extracted from the imaging data by measuring the distance travelled by the shock between two time delays, specifically between 300\,ps and 500\,ps. The uncertainty in the timing measurement is negligible due to the synchronization of the beams with an RMS $<$ 200\,fs. The uncertainty in the position is limited by the imaging resolution of $\approx500$\,nm. With this approach, the shock velocity results in $u_s = 14.3 \pm 1.3$\,km/s (at 42 $\mu m$ from the interaction point). The density reconstruction has been discussed in the previous section. The uncertainty on the pressure is calculated by propagating the uncertainties of the reconstructed density and shock velocity.

\end{appendices}


\bibliography{sn-bibliography}

\providecommand{\noopsort}[1]{}\providecommand{\singleletter}[1]{#1}%
\begin{thebibliography}{10}
\expandafter\ifx\csname url\endcsname\relax
  \def\url#1{\burl{#1}}\fi
\expandafter\ifx\csname urlprefix\endcsname\relax\def\urlprefix{URL }\fi
\providecommand{\bibinfo}[2]{#2}
\providecommand{\eprint}[2][]{\url{#2}}
\providecommand{\doi}[1]{\url{https://doi.org/#1}}
\bibcommenthead

\bibitem{Fowles}
\bibinfo{author}{Fowles, G.~R.} \emph{et~al.}
\newblock \bibinfo{title}{{Gas Gun for Impact Studies}}.
\newblock \emph{\bibinfo{journal}{Review of Scientific Instruments}} \textbf{\bibinfo{volume}{41}}, \bibinfo{pages}{984--996} (\bibinfo{year}{2003}).

\bibitem{Deeney}
\bibinfo{author}{Deeney, C.} \emph{et~al.}
\newblock \bibinfo{title}{Enhancement of x-ray power from a $\mathit{Z}$ pinch using nested-wire arrays}.
\newblock \emph{\bibinfo{journal}{Phys. Rev. Lett.}} \textbf{\bibinfo{volume}{81}}, \bibinfo{pages}{4883--4886} (\bibinfo{year}{1998}).

\bibitem{XHuang}
\bibinfo{author}{Huang, X.~B.} \emph{et~al.}
\newblock \bibinfo{title}{{Radiation characteristics and implosion dynamics of Z-pinch dynamic hohlraums performed on PTS facility}}.
\newblock \emph{\bibinfo{journal}{Physics of Plasmas}} \textbf{\bibinfo{volume}{24}}, \bibinfo{pages}{092704} (\bibinfo{year}{2017}).

\bibitem{Moses}
\bibinfo{author}{Moses, E.~I.}
\newblock \bibinfo{title}{Advances in inertial confinement fusion at the national ignition facility (nif)}.
\newblock \emph{\bibinfo{journal}{Fusion Engineering and Design}} \textbf{\bibinfo{volume}{85}}, \bibinfo{pages}{983--986} (\bibinfo{year}{2010}).
\newblock \bibinfo{note}{Proceedings of the Ninth International Symposium on Fusion Nuclear Technology}.

\bibitem{Spaeth}
\bibinfo{author}{Spaeth, M.~L.} \emph{et~al.}
\newblock \bibinfo{title}{Description of the nif laser}.
\newblock \emph{\bibinfo{journal}{Fusion Science and Technology}} \textbf{\bibinfo{volume}{69}}, \bibinfo{pages}{25--145} (\bibinfo{year}{2016}).

\bibitem{nora2015}
\bibinfo{author}{Nora, R.} \emph{et~al.}
\newblock \bibinfo{title}{Gigabar spherical shock generation on the omega laser}.
\newblock \emph{\bibinfo{journal}{Physical review letters}} \textbf{\bibinfo{volume}{114}}, \bibinfo{pages}{045001} (\bibinfo{year}{2015}).

\bibitem{boehly2011}
\bibinfo{author}{Boehly, T.} \emph{et~al.}
\newblock \bibinfo{title}{Velocity and timing of multiple spherically converging shock waves in liquid deuterium}.
\newblock \emph{\bibinfo{journal}{Physical Review Letters}} \textbf{\bibinfo{volume}{106}}, \bibinfo{pages}{195005} (\bibinfo{year}{2011}).

\bibitem{perez2022}
\bibinfo{author}{P{\'e}rez-Callejo, G.} \emph{et~al.}
\newblock \bibinfo{title}{Cylindrical implosion platform for the study of highly magnetized plasmas at laser megajoule}.
\newblock \emph{\bibinfo{journal}{Physical Review E}} \textbf{\bibinfo{volume}{106}}, \bibinfo{pages}{035206} (\bibinfo{year}{2022}).

\bibitem{Doeppner2018}
\bibinfo{author}{D\"oppner, T.} \emph{et~al.}
\newblock \bibinfo{title}{Absolute equation-of-state measurement for polystyrene from 25 to 60 mbar using a spherically converging shock wave}.
\newblock \emph{\bibinfo{journal}{Phys. Rev. Lett.}} \textbf{\bibinfo{volume}{121}}, \bibinfo{pages}{025001} (\bibinfo{year}{2018}).

\bibitem{KrausPhysRevE}
\bibinfo{author}{Kraus, D.} \emph{et~al.}
\newblock \bibinfo{title}{X-ray scattering measurements on imploding ch spheres at the national ignition facility}.
\newblock \emph{\bibinfo{journal}{Phys. Rev. E}} \textbf{\bibinfo{volume}{94}}, \bibinfo{pages}{011202} (\bibinfo{year}{2016}).

\bibitem{Kritcher_2016}
\bibinfo{author}{Kritcher, A.~L.} \emph{et~al.}
\newblock \bibinfo{title}{Shock hugoniot measurements of ch at gbar pressures at the nif}.
\newblock \emph{\bibinfo{journal}{Journal of Physics: Conference Series}} \textbf{\bibinfo{volume}{688}}, \bibinfo{pages}{012055} (\bibinfo{year}{2016}).

\bibitem{McBride2019}
\bibinfo{author}{McBride, E.~E.} \emph{et~al.}
\newblock \bibinfo{title}{Phase transition lowering in dynamically compressed silicon}.
\newblock \emph{\bibinfo{journal}{Nature Physics}} \textbf{\bibinfo{volume}{15}}, \bibinfo{pages}{89--94} (\bibinfo{year}{2019}).

\bibitem{Kraus2016}
\bibinfo{author}{Kraus, D.} \emph{et~al.}
\newblock \bibinfo{title}{Nanosecond formation of diamond and lonsdaleite by shock compression of graphite}.
\newblock \emph{\bibinfo{journal}{Nature Communications}} \textbf{\bibinfo{volume}{7}}, \bibinfo{pages}{10970} (\bibinfo{year}{2016}).

\bibitem{Briggs2017}
\bibinfo{author}{Briggs, R.} \emph{et~al.}
\newblock \bibinfo{title}{Ultrafast x-ray diffraction studies of the phase transitions and equation of state of scandium shock compressed to 82 gpa}.
\newblock \emph{\bibinfo{journal}{Phys. Rev. Lett.}} \textbf{\bibinfo{volume}{118}}, \bibinfo{pages}{025501} (\bibinfo{year}{2017}).

\bibitem{Hari_2023}
\bibinfo{author}{Hari, A.} \emph{et~al.}
\newblock \bibinfo{title}{High pressure phase transition and strength estimate in polycrystalline alumina during laser-driven shock compression}.
\newblock \emph{\bibinfo{journal}{Journal of Physics: Condensed Matter}} \textbf{\bibinfo{volume}{35}}, \bibinfo{pages}{094002} (\bibinfo{year}{2022}).

\bibitem{Kraus2017}
\bibinfo{author}{Kraus, D.} \emph{et~al.}
\newblock \bibinfo{title}{Formation of diamonds in laser-compressed hydrocarbons at planetary interior conditions}.
\newblock \emph{\bibinfo{journal}{Nature Astronomy}} \textbf{\bibinfo{volume}{1}}, \bibinfo{pages}{606--611} (\bibinfo{year}{2017}).

\bibitem{Hartley2018}
\bibinfo{author}{Hartley, N.~J.} \emph{et~al.}
\newblock \bibinfo{title}{Liquid structure of shock-compressed hydrocarbons at megabar pressures}.
\newblock \emph{\bibinfo{journal}{Phys. Rev. Lett.}} \textbf{\bibinfo{volume}{121}}, \bibinfo{pages}{245501} (\bibinfo{year}{2018}).

\bibitem{Gleason2022}
\bibinfo{author}{Gleason, A.~E.} \emph{et~al.}
\newblock \bibinfo{title}{Dynamic compression of water to conditions in ice giant interiors}.
\newblock \emph{\bibinfo{journal}{Scientific Reports}} \textbf{\bibinfo{volume}{12}}, \bibinfo{pages}{715} (\bibinfo{year}{2022}).

\bibitem{Luetgert2021}
\bibinfo{author}{L{\"u}tgert, J.} \emph{et~al.}
\newblock \bibinfo{title}{Measuring the structure and equation of state of polyethylene terephthalate at megabar pressures}.
\newblock \emph{\bibinfo{journal}{Scientific Reports}} \textbf{\bibinfo{volume}{11}}, \bibinfo{pages}{12883} (\bibinfo{year}{2021}).

\bibitem{Nagler:yi5007}
\bibinfo{author}{Nagler, B.} \emph{et~al.}
\newblock \bibinfo{title}{{The Matter in Extreme Conditions instrument at the Linac Coherent Light Source}}.
\newblock \emph{\bibinfo{journal}{Journal of Synchrotron Radiation}} \textbf{\bibinfo{volume}{22}}, \bibinfo{pages}{520--525} (\bibinfo{year}{2015}).

\bibitem{LasoGarcia2021}
\bibinfo{author}{Laso~Garcia, A.} \emph{et~al.}
\newblock \bibinfo{title}{Relax: the helmholtz international beamline for extreme fields high-intensity short-pulse laser driver for relativistic laser--matter interaction and strong-field science using the high energy density instrument at the european x-ray free electron laser facility}.
\newblock \emph{\bibinfo{journal}{High Power Laser Science and Engineering}} \textbf{\bibinfo{volume}{9}}, \bibinfo{pages}{e59} (\bibinfo{year}{2021}).

\bibitem{Yabuuchi2019}
\bibinfo{author}{Yabuuchi, T.} \emph{et~al.}
\newblock \bibinfo{title}{{An experimental platform using high-power, high-intensity optical lasers with the hard X-ray free-electron laser at SACLA}}.
\newblock \emph{\bibinfo{journal}{Journal of Synchrotron Radiation}} \textbf{\bibinfo{volume}{26}}, \bibinfo{pages}{585--594} (\bibinfo{year}{2019}).

\bibitem{Zastrau:ay5578}
\bibinfo{author}{Zastrau, U.} \emph{et~al.}
\newblock \bibinfo{title}{{The High Energy Density Scientific Instrument at the European XFEL}}.
\newblock \emph{\bibinfo{journal}{Journal of Synchrotron Radiation}} \textbf{\bibinfo{volume}{28}}, \bibinfo{pages}{1393--1416} (\bibinfo{year}{2021}).

\bibitem{CRLS}
\bibinfo{author}{Snigirev, A.}, \bibinfo{author}{Kohn, V.}, \bibinfo{author}{Snigireva, I.} \& \bibinfo{author}{Lengeler, B.}
\newblock \bibinfo{title}{A compound refractive lens for focusing high-energy x-rays}.
\newblock \emph{\bibinfo{journal}{Nature}} \textbf{\bibinfo{volume}{384}}, \bibinfo{pages}{49--51} (\bibinfo{year}{1996}).

\bibitem{Koch:98}
\bibinfo{author}{Koch, J.~A.} \emph{et~al.}
\newblock \bibinfo{title}{High-energy x-ray microscopy techniques for laser-fusion plasma research at the national ignition facility}.
\newblock \emph{\bibinfo{journal}{Appl. Opt.}} \textbf{\bibinfo{volume}{37}}, \bibinfo{pages}{1784--1795} (\bibinfo{year}{1998}).

\bibitem{Santos_2017}
\bibinfo{author}{Santos, J.~J.} \emph{et~al.}
\newblock \bibinfo{title}{Isochoric heating and strong blast wave formation driven by fast electrons in solid-density targets}.
\newblock \emph{\bibinfo{journal}{New Journal of Physics}} \textbf{\bibinfo{volume}{19}}, \bibinfo{pages}{103005} (\bibinfo{year}{2017}).

\bibitem{Quinn2009}
\bibinfo{author}{Quinn, K.} \emph{et~al.}
\newblock \bibinfo{title}{Laser-driven ultrafast field propagation on solid surfaces}.
\newblock \emph{\bibinfo{journal}{Phys. Rev. Lett.}} \textbf{\bibinfo{volume}{102}}, \bibinfo{pages}{194801} (\bibinfo{year}{2009}).

\bibitem{Bouffechoux10}
\bibinfo{author}{Buffechoux, S.} \emph{et~al.}
\newblock \bibinfo{title}{Hot electrons transverse refluxing in ultraintense laser-solid interactions}.
\newblock \emph{\bibinfo{journal}{Phys. Rev. Lett.}} \textbf{\bibinfo{volume}{105}}, \bibinfo{pages}{015005} (\bibinfo{year}{2010}).

\bibitem{Wilks01}
\bibinfo{author}{Wilks, S.~C.} \emph{et~al.}
\newblock \bibinfo{title}{{Energetic proton generation in ultra-intense laser--solid interactions}}.
\newblock \emph{\bibinfo{journal}{Physics of Plasmas}} \textbf{\bibinfo{volume}{8}}, \bibinfo{pages}{542--549} (\bibinfo{year}{2001}).

\bibitem{Borghesi2009}
\bibinfo{author}{Borghesi, M.} \emph{et~al.}
\newblock \bibinfo{title}{Relativistic current dynamics investigations by proton probing}.
\newblock \emph{\bibinfo{journal}{AIP Conference Proceedings}} \textbf{\bibinfo{volume}{1153}}, \bibinfo{pages}{319--330} (\bibinfo{year}{2009}).

\bibitem{10.1063/1.3262630}
\bibinfo{author}{Quinn, K.} \emph{et~al.}
\newblock \bibinfo{title}{{Modified proton radiography arrangement for the detection of ultrafast field fronts}}.
\newblock \emph{\bibinfo{journal}{Review of Scientific Instruments}} \textbf{\bibinfo{volume}{80}}, \bibinfo{pages}{113506} (\bibinfo{year}{2009}).

\bibitem{Borghesi2005a}
\bibinfo{author}{Borghesi, M.} \emph{et~al.}
\newblock \bibinfo{title}{High-intensity laser-plasma interaction studies employing laser-driven proton probes}.
\newblock \emph{\bibinfo{journal}{Laser and Particle Beams}} \textbf{\bibinfo{volume}{23}}, \bibinfo{pages}{291--295} (\bibinfo{year}{2005}).

\bibitem{Romagnani2005}
\bibinfo{author}{Romagnani, L.} \emph{et~al.}
\newblock \bibinfo{title}{Dynamics of electric fields driving the laser acceleration of multi-mev protons}.
\newblock \emph{\bibinfo{journal}{Physical Review Letters}} \textbf{\bibinfo{volume}{95}}, \bibinfo{pages}{195001} (\bibinfo{year}{2005}).

\bibitem{Toncian2006}
\bibinfo{author}{Toncian, T.} \emph{et~al.}
\newblock \bibinfo{title}{Ultrafast laser-driven microlens to focus and energy-select mega-electron volt protons}.
\newblock \emph{\bibinfo{journal}{Science}} \textbf{\bibinfo{volume}{312}}, \bibinfo{pages}{410--413} (\bibinfo{year}{2006}).

\bibitem{Quinn2012}
\bibinfo{author}{Quinn, K.} \emph{et~al.}
\newblock \bibinfo{title}{Weibel-induced filamentation during an ultrafast laser-driven plasma expansion}.
\newblock \emph{\bibinfo{journal}{Physical Review Letters}} \textbf{\bibinfo{volume}{108}}, \bibinfo{pages}{135001} (\bibinfo{year}{2012}).

\bibitem{yang2023dynamic}
\bibinfo{author}{Yang, L.} \emph{et~al.}
\newblock \bibinfo{title}{Dynamic convergent shock compression initiated by return current in high-intensity laser solid interactions}.
\newblock \emph{\bibinfo{journal}{arXiv preprint arXiv:2309.10626}}  (\bibinfo{year}{2023}).

\bibitem{Beg2004}
\bibinfo{author}{Beg, F.~N.} \emph{et~al.}
\newblock \bibinfo{title}{Return current and proton emission from short pulse laser interactions with wire targets}.
\newblock \emph{\bibinfo{journal}{Physics of Plasmas}} \textbf{\bibinfo{volume}{11}}, \bibinfo{pages}{2806--2813} (\bibinfo{year}{2004}).

\bibitem{johnson1994sesame}
\bibinfo{author}{Johnson, J.}
\newblock \bibinfo{title}{The sesame database}.
\newblock \bibinfo{type}{Tech. Rep.}, \bibinfo{institution}{Los Alamos National Lab.(LANL), Los Alamos, NM (United States)} (\bibinfo{year}{1994}).

\bibitem{fryxell2000flash}
\bibinfo{author}{Fryxell, B.} \emph{et~al.}
\newblock \bibinfo{title}{Flash: An adaptive mesh hydrodynamics code for modeling astrophysical thermonuclear flashes}.
\newblock \emph{\bibinfo{journal}{The Astrophysical Journal Supplement Series}} \textbf{\bibinfo{volume}{131}}, \bibinfo{pages}{273} (\bibinfo{year}{2000}).

\bibitem{dubey2009extensible}
\bibinfo{author}{Dubey, A.} \emph{et~al.}
\newblock \bibinfo{title}{Extensible component-based architecture for flash, a massively parallel, multiphysics simulation code}.
\newblock \emph{\bibinfo{journal}{Parallel Computing}} \textbf{\bibinfo{volume}{35}}, \bibinfo{pages}{512--522} (\bibinfo{year}{2009}).

\bibitem{Teague:82}
\bibinfo{author}{Teague, M.~R.}
\newblock \bibinfo{title}{Irradiance moments: their propagation and use for unique retrieval of phase}.
\newblock \emph{\bibinfo{journal}{J. Opt. Soc. Am.}} \textbf{\bibinfo{volume}{72}}, \bibinfo{pages}{1199--1209} (\bibinfo{year}{1982}).

\bibitem{tops}
\bibinfo{title}{Los alamos national laboratory- tops/atomic database}.
\newblock \urlprefix\url{https://aphysics2.lanl.gov/apps/}.

\bibitem{Paganin_2020}
\bibinfo{author}{Paganin, D.~M.} \emph{et~al.}
\newblock \bibinfo{title}{Boosting spatial resolution by incorporating periodic boundary conditions into single-distance hard-x-ray phase retrieval}.
\newblock \emph{\bibinfo{journal}{Journal of Optics}} \textbf{\bibinfo{volume}{22}}, \bibinfo{pages}{115607} (\bibinfo{year}{2020}).

\bibitem{Mitchell}
\bibinfo{author}{Mitchell, A.~C.} \emph{et~al.}
\newblock \bibinfo{title}{{Equation of state of Al, Cu, Mo, and Pb at shock pressures up to 2.4 TPa (24 Mbar)}}.
\newblock \emph{\bibinfo{journal}{Journal of Applied Physics}} \textbf{\bibinfo{volume}{69}}, \bibinfo{pages}{2981--2986} (\bibinfo{year}{1991}).

\bibitem{Ragan}
\bibinfo{author}{Ragan, C.~E.}
\newblock \bibinfo{title}{Shock-wave experiments at threefold compression}.
\newblock \emph{\bibinfo{journal}{Phys. Rev. A}} \textbf{\bibinfo{volume}{29}}, \bibinfo{pages}{1391--1402} (\bibinfo{year}{1984}).

\bibitem{planetpaper}
\bibinfo{author}{Smith, R.~F.} \emph{et~al.}
\newblock \bibinfo{title}{Equation of state of iron under core conditions of large rocky exoplanets}.
\newblock \emph{\bibinfo{journal}{Nature Astronomy}} \textbf{\bibinfo{volume}{2}}, \bibinfo{pages}{452--458} (\bibinfo{year}{2018}).

\bibitem{Kritcher2020}
\bibinfo{author}{Kritcher, A.~L.} \emph{et~al.}
\newblock \bibinfo{title}{A measurement of the equation of state of carbon envelopes of white dwarfs}.
\newblock \emph{\bibinfo{journal}{Nature}} \textbf{\bibinfo{volume}{584}}, \bibinfo{pages}{51--54} (\bibinfo{year}{2020}).

\bibitem{data}
\bibinfo{title}{Euxfel data repository - hed 4597} (\bibinfo{year}{2023}).
\newblock \urlprefix\url{https://doi.org/10.22003/XFEL.EU-DATA-004597-00}.

\bibitem{Preston_2020}
\bibinfo{author}{Preston, T.} \emph{et~al.}
\newblock \bibinfo{title}{Design and performance characterisation of the hapg von hámos spectrometer at the high energy density instrument of the european xfel}.
\newblock \emph{\bibinfo{journal}{Journal of Instrumentation}} \textbf{\bibinfo{volume}{15}}, \bibinfo{pages}{P11033} (\bibinfo{year}{2020}).

\bibitem{Kirkwood2019}
\bibinfo{author}{Kirkwood, H.~J.} \emph{et~al.}
\newblock \bibinfo{title}{Initial observations of the femtosecond timing jitter at the european xfel}.
\newblock \emph{\bibinfo{journal}{Opt. Lett.}} \textbf{\bibinfo{volume}{44}}, \bibinfo{pages}{1650--1653} (\bibinfo{year}{2019}).

\end{thebibliography}

\end{document}